\documentclass[11pt,BCOR2mm,DIV12]{scrartcl}

\newcommand{\xmin}{x_{\min}}
\newcommand{\bmtheta}{\mbox{\boldmath$\theta$}}
\newcommand{\bmphi}{\mbox{\boldmath$\phi$}}

\usepackage{graphicx}
\usepackage{amssymb,amsfonts,amsmath,mathtools}
\usepackage{booktabs}
\usepackage{lipsum}

\graphicspath{{graphics/}}

\begin{document}

\title{A complete data frame work for fitting power law distributions}
\author{Colin S. Gillespie}
\maketitle

\begin{abstract}
  Over the last few decades power law distributions have been suggested as
  forming generative mechanisms in a variety of disparate fields, such as,
  astrophysics, criminology and database curation. However, fitting these heavy
  tailed distributions requires care, especially since the power law behaviour
  may only be present in the distributional tail. Current state of the art
  methods for fitting these models rely on estimating the cut-off parameter
  $\xmin$. This results in the majority of collected data being discarded. This
  paper provides an alternative, principled approached for fitting heavy tailed
  distributions. By directly modelling the deviation from the power law
  distribution, we can fit and compare a variety of competing models in a single
  unified framework.
\end{abstract}

\section{Introduction} \label{s:intro}

Power law probability distributions have the relatively simple form of
\begin{equation}\label{1}
p(x) \propto x^{-\alpha} 
\end{equation}
where $\alpha > 1$ and $x >0$. The parameter $\alpha$, is often referred to as
the \textit{exponent} or \textit{scaling} parameter. Although straightforward,
these distributions have gathered scientific interest from many areas, including
terrorism, astrophysics, neuroscience, biology, database curation and
criminology\cite{Clauset2007a,Michel2011,Beggs2003,Yu2008,Bell2012, Duijn2014}.

This apparent ubiquity of power laws in a wide range of disciplines was
questioned by Stumpf and Porter\cite{Stumpf2012}. The authors' point out that
many ``observed'' power law relationships are highly suspect. In particular,
estimating the power law exponent on a log-log plot, whilst appealing, is a very
poor technique for fitting these types of models. Instead, a systematic,
principled and statistical rigorous approach should be applied.

Power law distributions are often described as ``scale-free'' - indicating that
common, small events are qualitatively similar to large, rare events.
Identifying a power law can highlight the presence of underlying generative
mechanism of interest.

Determining whether a quantity follows a power law distribution is complicated
by the large fluctuations in the tail of the empirical distribution. These large
spikes follow naturally from the power law distribution. For the continuous
power law distribution, the raw moments are
\[
E[X^m] = \int_{\xmin}^{\infty} x^m p(x)\, dx = \frac{\alpha - 1}{\alpha - 1 -m}
\xmin^m \;.
\]
So when 
\begin{itemize}
\item $1< \alpha \le 2$, all moments diverge, i.e. $E[X] = \infty$;
\item $2 < \alpha \le 3$, all second and higher-order moments diverge, i.e. $E[X^2] = \infty$;
\item $3 < \alpha \le m+1$, all $m^{\text{th}}$ and higher-order moments diverge, i.e.
  $E[X^m] = \infty$.
\end{itemize}
However, large outlying values in the tail of the distribution are not unique to
power laws. Many other ``standard'' distributions, such as the log normal, are
characterised with heavy tails.

A further complication, is that the power law distribution may only be
appropriate in the distributional tail, i.e. power law patterns only occur when
$x \ge \xmin$. While the value of the $\xmin$ cut-off could be estimated by eye
using log-log plots, this would obviously be a poor inference technique.

Clauset \textit{et al}, 2009 introduced a principled set of methods for fitting
and testing power law distributions\cite{Clauset2009}. Their approach is
straightforward and appealing. They couple a distance-based test for estimating
$\xmin$, with a standard maximum likelihood technique for inferring $\alpha$.
Competing models can be compared using a likelihood ratio test\cite{Vuong1989}.
However, their method does have three main draw-backs. First, by fitting $\xmin$
we are \textit{discarding} all data below that cut-off. Second, it is unclear
how to compare distributions where each distribution has a different $\xmin$.
Third, although it is possible to make predictions in the tail of the
distribution\cite{Clauset2013}, making future predictions over the entire data
space is not possible since values less than $\xmin$ have not been directly
modelled.

In a recent paper, Peterson \textit{et al.} 2013, propose a generative mechanism
that describes the formation of heavy tailed distributions\cite{Peterson2013}.
This neat formulation uses a statistical physics framework to express the
underlying model in terms of shared costs and economies of scale. While this
formulation fits the entire data set, some of the fits in the tail of the
distributions were not optimal (in particular, the Github and Petster data sets
shown in figure \ref{F1}a).

Estimating $\xmin$ directly can impose a strange dichotomy between relevant and
irrelevant observations. Instead, we adopt a different approach. Rather than
directly estimating $\xmin$ and thereby discarding data, we model the entire
data set as the \textit{deviation} away from a power law (or other heavy tailed)
distribution. By modelling the entire dataset, a number of standard statistical
techniques, which are not straightforward in other power law modelling
frameworks, become amenable. For example,
\begin{itemize}
\item prediction of future values of the phenomena of interest (see section \ref{E2});
\item comparing different distributions models using AIC and BIC (see section \ref{E3});
\item investigating model fit (see sections \ref{E1}, \ref{E2} and \ref{E3});  
\item comparing different datasets (see section \ref{E2}).
\end{itemize}

\section{Method}

In this paper we propose to model heavy tailed distributions using the distribution
\begin{align}\label{2}
f(x; \bmtheta, \bmphi) &= Pr(X=x) \nonumber \\
&= \frac{g(x; \bmtheta) \times D(x; \bmphi)}{C(\bmtheta, \bmphi)}
\quad \text{for} \quad x=1, \ldots
\end{align}
where $(\bmtheta, \bmphi)$ are model parameters and $C(\bmtheta, \bmphi)$ is a
normalising constant. The function $f(\cdot)$ consists of two key components.
\begin{itemize}
\item A heavy-tailed distribution: $g(x; \bmtheta)$. This could be, for example, a
  power law, log normal or other heavy tailed distribution. Typically, this
  function would fit the tail of the distribution well.
\item A difference function: $D(x; \bmphi)$. This function describes the deviation
  between the heavy tailed distribution, $g(x; \bmtheta)$ and the data. The key
  properties are $D(x; \bmphi) > 0$ and as $x \rightarrow \infty$, $D(x; \bmphi)
  \rightarrow 1$.
\end{itemize}
Typical functional forms for $D(x; \bmphi)$ are
\[
D(x; \bmphi) = 1-e^{-(\phi_0 + \phi_1 (x-1))} \quad \text{Unit exponential CDF}
\]
and
\[
D(x; \bmphi) = \frac{1}{1+e^{-(\phi_0 + \phi_1 (x-1))}} \quad \text{Inverse logistic function.}
\]
To fit model (\ref{2}) we follow the Bayesian paradigm. Let $\pi(\bmtheta)$ and
$\pi(\bmphi)$ denote the respective prior densities for $\bmtheta$ and $\bmphi$,
and $\mathcal{D}$ is the observed data. Fully Bayesian inference can then
proceed by sampling
\begin{equation}
\pi(\bmtheta, \bmphi \,\vert\, \mathcal{D}) \propto \pi(\bmtheta) \pi(\bmphi)
f(\mathcal{D} \,\vert\, \bmtheta, \bmphi) \;,
\end{equation}
using a Markov chain Monte Carlo algorithm. To explore the parameter space, a
multivariate Gaussian random walk proposal can be used, with the tuning parameter
obtained from pilot runs.

\subsection{Parameter estimation}

Inferences regarding the parameters of $g(x; \mathbf{\bmtheta})$, e.g. the power
law scaling parameter, are obtained from $\pi(\bmtheta, \bmphi \,\vert\,
\mathcal{D})$. Intuitively, the cut-off parameter $\xmin$ corresponds to when
$D(x; \bmphi) \simeq 1$. So a marginal posterior density for $\xmin$ can be
obtained by calculating
\begin{equation}
\arg \min_x D(x; \bmphi) > \alpha \;,
\end{equation}
where $\bmphi$ are posterior samples and $\alpha$ is a threshold.

\begin{figure*}[!t]
  \centering
  \includegraphics[width=\textwidth]{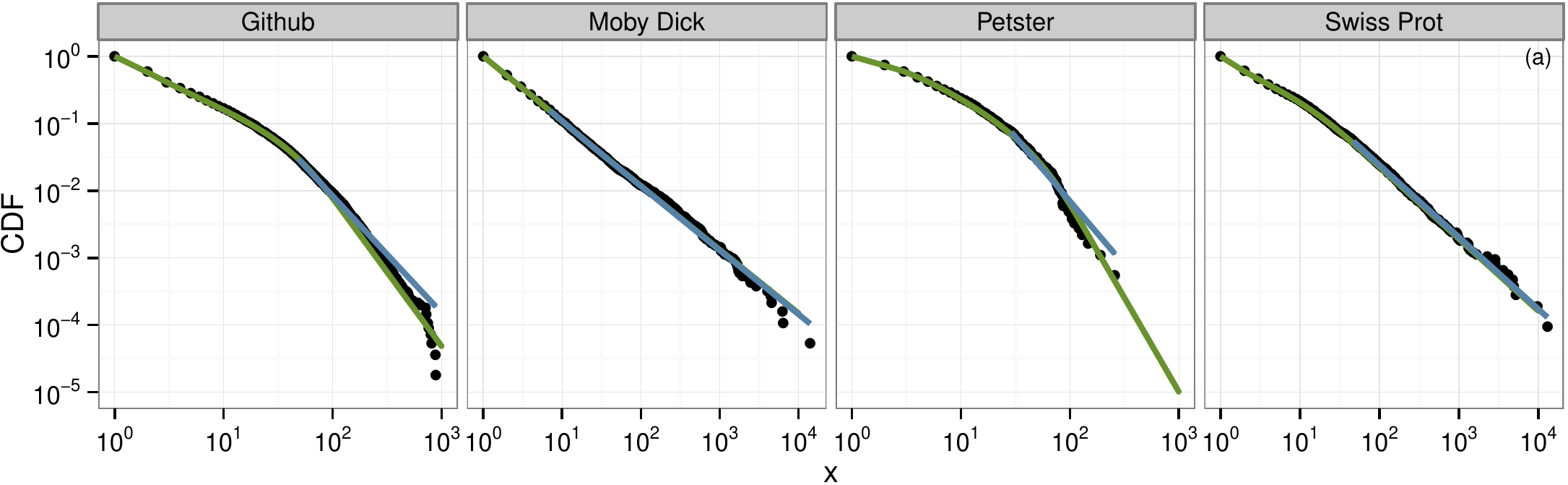}
  \includegraphics[width=\textwidth]{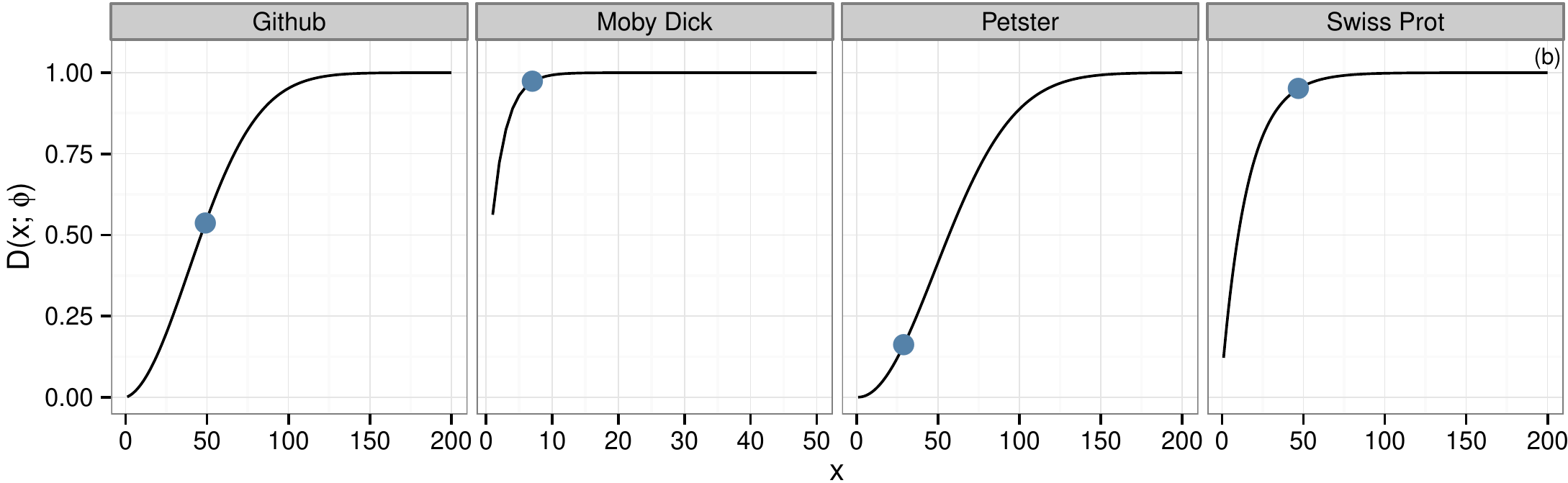}
  \includegraphics[width=\textwidth]{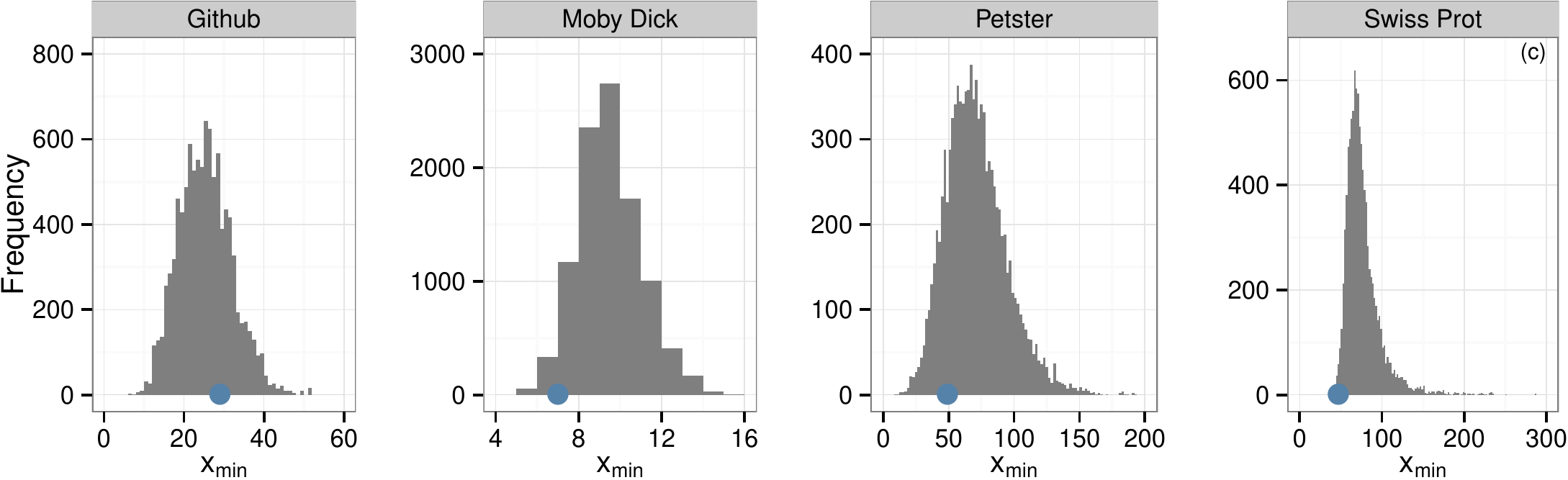}
  \caption{(a) Empirical distributions for the four example data sets. The green
    is the fitted function $f(x; \theta, \bmphi)$. The blue line is a fitted
    power law function, with parameters estimated using the CSN method. (b) A
    plot of the difference function $D(x; \bmphi)$ (where $\bmphi$ has been set
    to their posterior mean values). The blue dots shows the estimated value of
    $\xmin$ using the CSN method. (c) Estimated values of $\xmin$, obtained via
    $\arg \min_x D(x; \bmphi) > 0.95$.} \label{F1}
\end{figure*}

\section{Examples}\label{E1}

Figure \ref{F1} gives four illustrative examples.
\begin{itemize}
\item Project membership on the social coding web site GitHub\cite{Chacon09}.
\item Occurrences of unique words in the novel Moby Dick\cite{Newman2005}.
\item Friendships between users of the Petster social networking site
  Hamsterster\cite{Kunegis2013}.
\item Occurrences of words in the Swiss-Prot database\cite{Bell2012}.
\end{itemize} 
Each of the distributions in figure \ref{F1}a has a ``long-tail''. However, the
distributional shapes differ. The word occurrence data sets - Moby Dick and
Swiss Prot - have long power law like distributions. Whereas the social
networking data sets - Petster and Github - have a more curved distribution. 

In each example $g(x; \bmtheta)$ is a power law distribution (with $\xmin=1$),
i.e.
\begin{equation}\label{3}
g(x, \theta) = \frac{x^{-\theta}}{\zeta(\theta)} \quad \text{for} \quad x = 1, \ldots
\end{equation}
where $\theta$ is the power law scaling parameter and
\[
\zeta(\theta) = \sum_{i=1}^{\infty} \frac{1}{i^{\theta}}
\]
is Riemann zeta function. To model the deviation away from the power law
distribution, we use the unit exponential cumulative density function
\begin{equation}\label{4}
D(x; \bmphi) = 1-e^{-\phi_0- \phi_1 (x-1)- \phi_2 (x-1)^2}
\end{equation}
where $\bmphi = \{\phi_0, \phi_1, \phi_2\} \ge 0$. So $D(x=1; \bmphi) = 1-e^{-\phi_0}$ and $g(x =
\infty; \bmphi) = 1$. 

Figure \ref{F1}a gives the empirical CDF of each data set and the fitted
function $f(x; \theta, \bmphi)$. Also shown is a fitted power law distribution
where $\xmin$ and the scaling parameter $\theta$ were estimated using the CSN
method from Clauset \textit{et al.}\cite{Clauset2009}. Figure \ref{F1}b plots
the difference function $D(x; \bmphi)$ along with the estimated $\xmin$ value
obtained from the CSN method.

For each data set in figure \ref{F1}a, $f(x; \theta, \bmphi)$ provides an
excellent fit. However, the key benefit is that all data is used (see table
\ref{T1} for an exact breakdown). For example, in the Moby Dick data set since
$\xmin=7$, fitting a power law using CSN would result in discarding around 84\%
of the collected data.

For the network data sets, the standard power law fit is relatively poor,
however, our flexible formulation still provides an excellent fit.

Inference can also be made about plausible values $\xmin$ by calculating 
\[
\arg \min_x D(x; \bmphi) > 0.95 \;,
\]
where $\bmphi$ are samples from the posterior. The results given in figure
\ref{F1} are broadly consistent with the CSN method. However, the difference
method also provides credible regions to assess parameter uncertainty.

\begin{table}[t]
\centering
\caption{Summary statistics from model fits relating to figure \ref{F1}. The
  values of $\hat{x}_{\text{min}}$ and $\hat \alpha$ were obtained CSN method. The
  parameter $\bar \theta$ is the average posterior value of $\theta$ in expression
  \ref{3}. The column, \% $< \hat{x}_{\text{min}}$, gives the proportion of data that
  are less the $\hat{x}_{\text{min}}$.}\label{T1}
\begin{tabular}{@{} l llll @{}}
\toprule
Data set & $\hat{x}_{\text{min}}$ & \% $< \hat{x}_{\text{min}}$ & $\hat \alpha$ & $\bar \theta$ \\
\midrule
Github & 49 & 97 & 2.74 & 3.21 \\
Moby Dick & \phantom{0}7  & 84 & 1.95 & 1.95\\
Petster &  29 & 93 & 2.90  & 3.80 \\
Swiss Prot & 47  & 95 & 2.07 & 2.07\\
\bottomrule
\end{tabular}
\end{table}

\subsection{Data set comparison}\label{E2}
\begin{figure}[!t]
\centering
\includegraphics[width=0.5\textwidth]{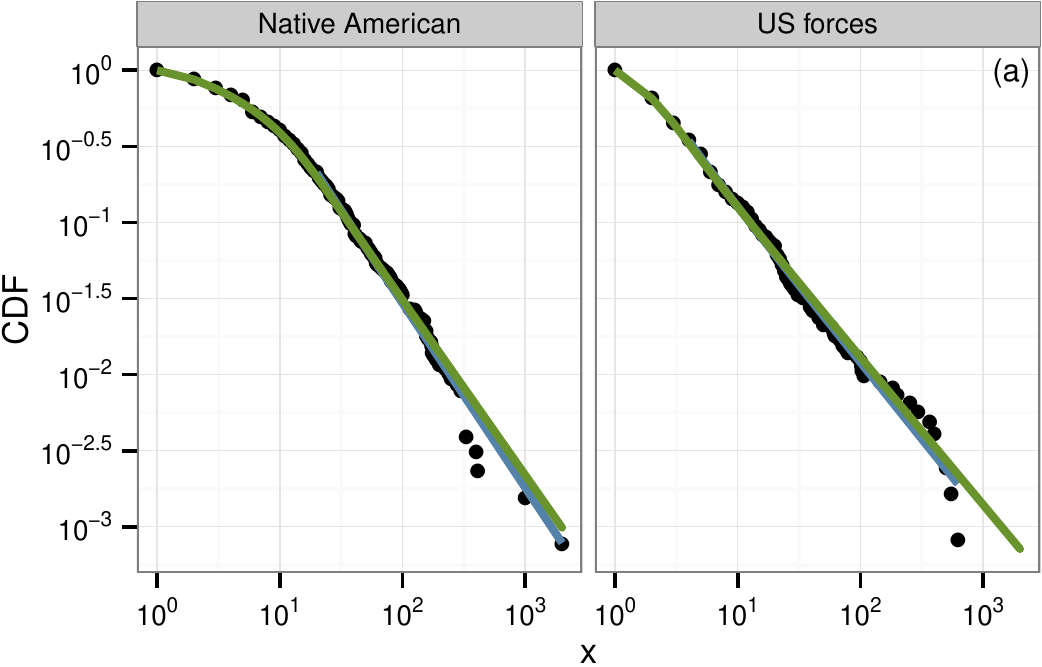}%
\includegraphics[width=0.5\textwidth]{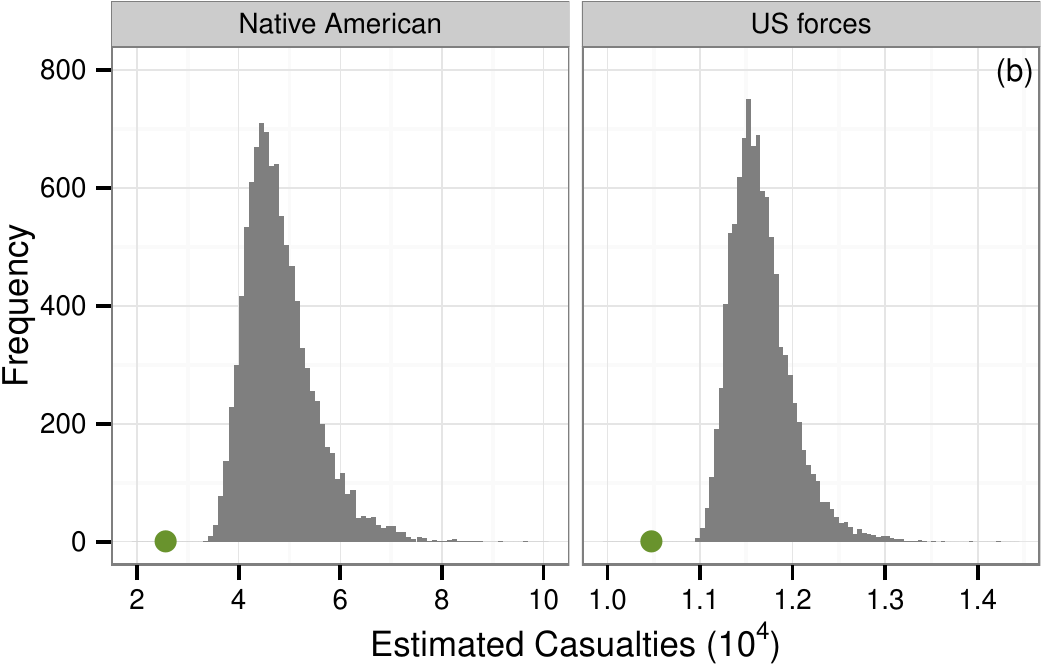}
\includegraphics[width=\textwidth]{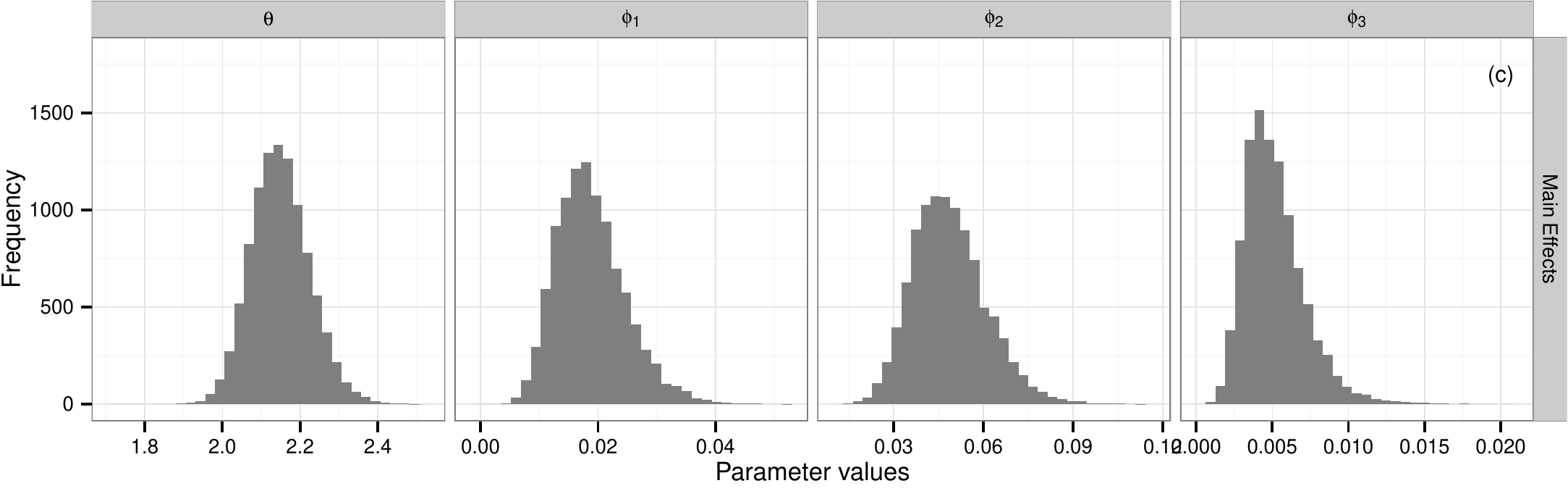}
\includegraphics[width=\textwidth]{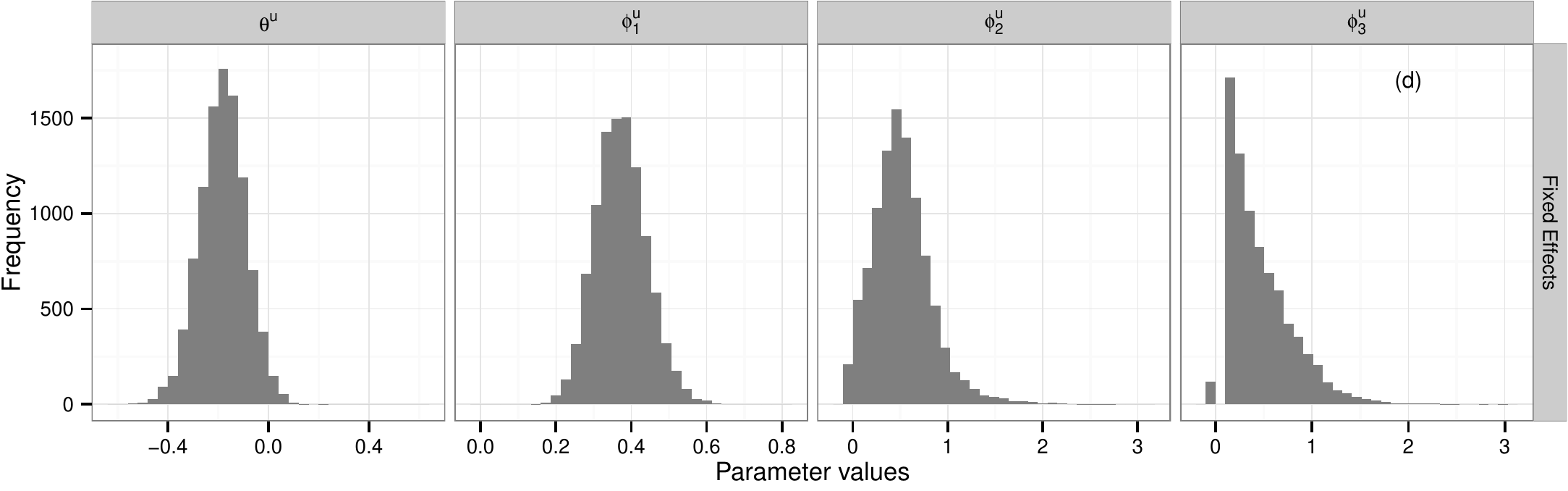}
\caption{Casualty rates from the American Indian Wars. (a) Empirical CDFs with
  lines of best fit. Each data point represents the number of casualties at a
  recorded battle. The blue line is the fitted distribution obtained from the
  CSN method. The green line is the difference method.(b) The inferred total
  number of casualties. The green dot shows the observed number of casualties.
  (c) \& (d) Posterior distributions for the parameters in $D(x;
  \cdot)$.}\label{F2}
\end{figure}

Recently Friedman investigated casualty rates in the American Indian Wars, 1776
to 1890\cite{Friedman2013}. Casualty rates are often controversial, since any
estimation will involve inferences about missing data. It has been observed
that many violent events, ranging from homicides to world wars, follow a
power law distribution. For example, power laws have been used to characterise
terrorist attacks\cite{Clauset2007}, inter-state wars\cite{Cederman2003} and
insurgent attacks\cite{Bohorquez2009}. Friedman uses this power law insight to
fit distributions to the American Indian conflict, then infer missing casualties.

The casualty numbers for the Native American and US forces are shown in figure
\ref{F2}a. The Native American casualties were modelled using
\begin{align*}
g(x; \theta) &= \frac{x^{-\theta}}{\zeta(\theta)}\\ \nonumber
h(x; \bmphi) &= 1-e^{-\phi_0- \phi_1 (x-1)- \phi_2 (x-1)^2}
\end{align*}
and the US forces casualties as
\begin{align*}
g_U(x; \theta^*) &= \frac{x^{-\theta^*}}{\zeta(\theta^*)}\\ \nonumber
h_U(x; \bmphi^*) &= 1-e^{-\phi_0^*- \phi_1^* (x-1)- \phi_2^* (x-1)^2}
\end{align*}
where
\[
\theta^* = \theta + \theta^u \quad \text{and} \quad \phi_i^* = \phi_i + \phi_i^u \;.
\]
Hence the parameters $\theta^u$ and $\bmphi^u$ directly model the difference
between the US forces and Native American casualties. The green line in
figure \ref{F2}a gives fitted function and blue line gives the standard power
law fit using CSN method.

Assuming that the true underlying distribution is a power law, then the function
$D(x; \bmphi)$ describes the number of missing casualties in the data. To infer
the total number of casualties, we integrate the posterior over the uncertainty
associated with the parameter. In other words, this predictive distribution is
taken to be the posterior average of realisations from $g(x; \theta)$ and
$g_U(x; \theta^*)$. This integration yields figure \ref{F2}b. The predictions
given in figure \ref{F2}b are consistent with Friedman's analysis.

An additional benefit of fitting models in this way, is that it is
straightforward to compare data sets. In this example, Figure \ref{F2}c and
\ref{F2}d give the posterior distribution for the parameters. In general, the
posterior of $\theta^u$ is negative, indicating that the causality rate is lower
for US forces. Furthermore, since $D(x;\bmphi + \bmphi^u) \ge D(x;\bmphi^u)$ the
reporting of US force casualties was more consistent. Again, this agrees with
the observations of Friedman.

\subsection{Model comparison}\label{E3}

\begin{figure}[t]
\centering
\includegraphics[width=0.5\textwidth]{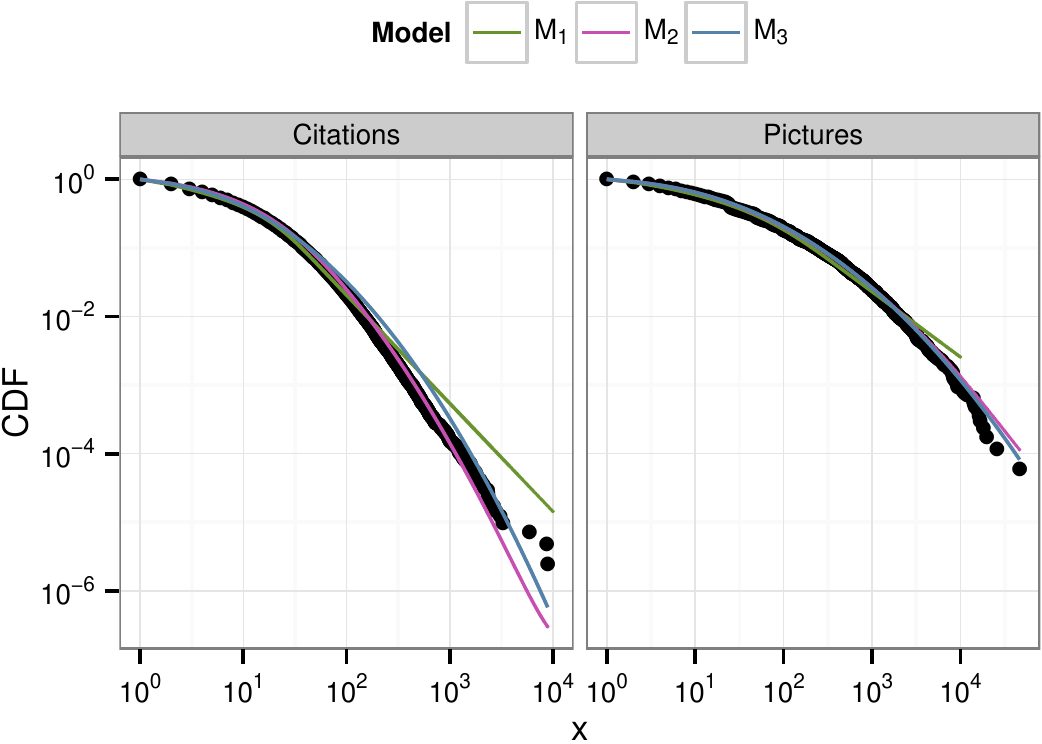}%
\caption{Citation and pictures data sets (details given in the text). The
  empirical CDF is given along with lines of best fit for models (i) $M_1$ -
  power law with unit exponential difference function. (ii) $M_2$ - discrete log
  normal with unit exponential difference function. (iii) $M_3$ - discrete log
  normal.}\label{F3}
\end{figure}
By fitting models to the entire data set, existing model comparison techniques
can be leveraged. Figure \ref{F3} show two data sets.
\begin{itemize}
\item The number of citations received between publication and June 1997 by
  scientific papers published in 1981 and listed in the Science Citation
  Index\cite{Redner1998};
\item The connections in the bipartite picture tagging network of
  vi.sualize.us.\cite{Kunegis2013}.
\end{itemize}
For each data set we fitted three models.
\begin{enumerate}
\item $M_1$: where the deviation from a power law was modelled using a unit
  exponential function, i.e. expressions \ref{3} and \ref{4}.
\item $M_2$: where the deviation from a discrete log normal was modelled using a
  unit exponential function.
\item $M_3$: a discrete log normal distribution.
\end{enumerate}
Figure \ref{F3} show that the log normal based models, $M_2$ and $M_3$, fit the
data set reasonably well. To formally compare models, we calculated the Bayesian
Information Criterion (BIC) using the posterior means (see table \ref{T2}). In
both examples, the power law model was rejected when compared to log normal
based models.
\begin{table}[t]
\centering
\caption{Bayesian information criterion (BIC) values for competing model
  in the citation and pictures data set shown in figure \ref{F3}. The smallest
  BIC value for each model is shown in bold.}\label{T2}
\begin{tabular}{@{} l lll @{}}
\toprule
&  \multicolumn{3}{c}{Model}\\
\cmidrule(l){2-4}
Data set & $M_1$ & $M_2$ & $M_3$ \\
\midrule
Citations & $2,919,701$ & $\mathbf{2,917,967}$ & $2,921,903$\\
Pictures & $\phantom{2,}167,641$ & $\phantom{2,}167,513$ & $\mathbf{\phantom{2,}167,490}$ \\
\bottomrule

\end{tabular}
\end{table}

\section{Discussion}

Typically, researchers suggest that the power law feature is only present in the
distributional tail. By modelling the deviation away from the power law (or
other heavy tailed distribution), we have created a flexible and general
framework. We reiterate that by modelling the \textit{entire} data set, we
circumvent the problem of estimating $\xmin$, and thereby discarding part of the
data. Since we have avoided the $\xmin$ issue, standard statistical tools
automatically become available.

By adopting a Bayesian framework, more complex observed data structures can be
incorporated. For example, Virkar and Clauset recently consider ``binned'' data
sets\cite{Virkar2014}. This relates to a number of data sets where the
observations have been rounded. In the framework proposed in this paper, we
could simply introduce an appropriate data error model, and integrate out the
uncertainty using Markov chain Monte Carlo and the analysis would proceed as
before.

\section*{Computing details}

All simulations were performed on a machine with 4GB of RAM and with an Intel
quad-core CPU using R\cite{RCoreTeam2013}. The CSN power law fits were obtained
using the \texttt{poweRlaw} package\cite{Gillespie2014}. All code associated
with this paper can be obtained from
\begin{center}
https://github.com/csgillespie/plmcmc
\end{center}

\section*{Acknowledgements}

We would like to thank Aaron Clauset and Jeff Friedman for their helpful
comments on this manuscript.

\bibliographystyle{plain}
\bibliography{refs}

\end{document}